\documentclass[12pt]{article}
\usepackage{color}
\usepackage{latexsym}
\usepackage{amsmath}
\usepackage{amsfonts}
\usepackage{amssymb}
\usepackage{indentfirst} 
\numberwithin{equation}{section}
\newcommand{\be}{\begin{equation}}
\newcommand{\ee}{\end{equation}}
 
\newcommand{\mR}{{\mathbb R}}

\newcommand{\mA}{{\mathcal A}}
\newcommand{\mW}{{\mathcal W}}

\title{Memory effect, conformal symmetry and  gravitational plane  waves}
\author{K. Andrzejewski\footnote{Corresponding author, e-mail: k-andrzejewski@uni.lodz.pl},\quad  S. Prencel
\vspace*{0.8cm}
\\
\small  Department of  Computer Science,  Faculty of Physics    and Applied Informatics, \\ \small
 University of  Lodz, Pomorska 149/153,
90-236, Lodz, Poland\\
}
\date{}
\begin{document}
\maketitle 
\begin{abstract}
We  discuss in some detail  the interaction of classical particles, including the  scattering and  memory effect,   with  a pulse of  gravitational plane wave. The key point is the  conformal symmetry of  gravitational  plane waves.  In particular,   we  obtain, in the limit of short pulse,  some results for  impulsive gravitational waves.  Furthermore, in the general case, we give certain conditions which allow us   to  completely describe  the interaction in terms of the singular  Baldwin-Jeffery-Rosen   coordinates. 
\end{abstract}
\section{Introduction}
Gravitational  waves have been for many years a matter of intensive study and controversy  \cite{b1a}-\cite{b1f}.
Recently they gained a new interest both for experimental and theoretical reasons. Although indirect evidence for their existence  had been obtained from the observation of binary pulsar system PSR 1913+16 \cite{b1g}
only recently the direct observation of gravitational waves from a pair of  merging black holes \cite{b1h} and binary neutron star inspiral \cite{b1i} have been possible.  From the theoretical point of view a new interesting set of ideas relating to  asymptotic symmetries soft theorems and gravitational memory effect has recently emerged  \cite{b2a}-\cite{b2g}. 
The gravitational memory effect consists, roughly speaking, in the change in separation of freely falling particles after the passage of  short burst  of gravitational wave  \cite{b1e}-\cite{b3e}. On the other hand, 
soft graviton theorems are related to gauge transformations (diffeomorphisms)   of the asymptotically flat spacetime (which do not  tend to identity at the infinity) \cite{b4a,b4b}. Both the issues  are   related to geodesic (deviation) equations and asymptotically flat metrics. 
In many cases one can restrict the considerations to plane waves. This is motivated by the fact  that  far from the source one can approximate the gravitational wave in a neighborhood  of  detector  by  an exact  plane wave  (assuming that  the back reaction of detector is negligible).  Especially interesting is the case of the  so called  impulsive gravitational  waves \cite{b5a}-\cite{b5i}.  
\par In view  of the above remarks it would be interesting to prove analytically solvable examples of  gravitational plane waves.  Taking into account  that higher symmetry of the system is the more likely  it is analytically solvable we consider the  wave profile carrying the maximal conformal symmetry, allowed for non-flat  plane waves.  Such a choice allows us to  provide the explicit   description  of the behaviour of  test particles  and study of memory effect  as well as classical cross section.  Moreover, by taking an appropriate limit  we give explicit description, including the form of conformal symmetry generators,   of Dirac delta profile. 
\par 
Moreover, we study also in some detail  the problem of  complete description  of gravitational plane waves in Baldwin-Jeffery-Rosen  (BJR) coordinates. It  is well know that the BJR map does not  cover the whole spacetime manifold corresponding to the  gravitational plane  wave. On the other hand,  BJR coordinates  seem to be some importance  in    understanding   inequivalent ground states (vacua) \cite{b2c,b2d}. 
Therefore, a deeper insight into the structure  of BJR coordinates would be profitable.
\section{Conformal symmetry and geodesics}
\label{s1}
\subsection{Preliminaries}
In relativity, a Ricci  flat plane  wave  is called a  gravitational plane wave, or an  exact gravitational wave. In the so called   Brinkmann (B)  coordinates \cite{b1a} it is described by the metric tensor 
\be
\label{e000}
g=\sum_{\sigma,\sigma'}K_{\sigma\sigma'}(u)X^\sigma X^{\sigma'}du^2+2dudV+\sum_{\sigma}(dX^\sigma)^2,
\ee
where indices $\sigma,\sigma'=\pm$ and
\be 
\sum_{\sigma,\sigma'}K_{\sigma\sigma'}(u)X^\sigma X^{\sigma'}=\frac 12\mA_+(u)((X^+)^2-(X^-)^2)+\mA_\times(u) X^+X^-.
\ee
To begin  with,  one can consider the linear  polarization,   i.e., $\mA_\times\equiv 0$. Denoting $\mA_+(u)\equiv \mA(u)$,  we get the   metric    
\be
\label{e00}
g=\frac 12\mA ((X^+)^2-(X^-)^2){du}^2+2{du}{}{dV}{} +\left({dX^+ }{ }\right)^2+\left({d X^-}{}\right)^2 ,
\ee
and the Lagrangian of the test  particle
\be
L=m\left[\frac 12\mA ((X^+)^2-(X^-)^2)\left(\frac{du}{d \tau}\right)^2+2\frac{du}{d\tau}\frac{dV}{d\tau} +\left(\frac{dX^+ }{ d \tau}\right)^2+\left(\frac{d X^-}{d\tau}\right)^2\right].
\ee
Let us define
\be
\label{e14} 
P^\mu\equiv m\frac{d X^\mu}{d \tau}=\frac{d X^\mu}{d \lambda},
\ee
 where  $\lambda =\frac {\tau}{m}$. Then 
\be
\label{e0}
\textrm{const}=P_V\equiv g_{V\mu}\frac{dX^\mu}{d\lambda}=P^u=\frac{du}{d\lambda}.
\ee
Thus 
\be
\label{e9}
u=\lambda P_V.
\ee
Then the geodesic equations are of the form\footnote{If not otherwise stated, we assume that $\mA$ is a continuous function; consequently   the geodesics are defined on the whole real line.    } 
\be
\label{e9a}
\frac{d^2 X^\sigma}{d u ^2}=  \frac{\sigma\mA}{2}X^\sigma, \quad \sigma=\pm,
\ee
and
\be
\label{e9b}
\frac{d^2 V}{du^2}+\sum_\sigma\sigma X^\sigma\left(\mA \frac{d X^\sigma}{d u}+\frac 14\frac {d \mA }{d u}X^\sigma \right)=0.
\ee
The observable quantities are connected with the geodesic deviation equations (see \cite{b2d} for resume); in the case of    the transverse-space coordinates $X^\sigma$ the latter  coincide  with eqs. (\ref{e9a}). Thus the problem of the   interaction of the particle with the gravitational waves or the scattering problem can be reduced to the analysis of  solutions to eqs. (\ref{e9a}) and (\ref{e9b}). Although, in general, this set of equations  cannot be explicitly solved,  there are some special cases where the solution  is possible; they are  mainly related to the symmetry of the metric; of course the most interesting  cases are the ones with the maximal symmetry.
\par
It is well known that the generic dimension of the  isometry group of the gravitational plane waves   is  five \cite{b1e,b6a,b6b}. However, when $\mA_1 =\textrm{const}$ or $\mA_2 =\frac {\textrm{const}}{u^2}$ (and  a  suitable generalization   to circular polarization, see e.g.,  \cite{b1e})  the dimension of the isometry group of $g$  is  six and then  one can  explicitly solve the geodesic equations. The first  case can be used, for example,  to  describe  the  gravitational wave  which   is  a sandwich  between two Minkowskian regions;  the second one is not  geodesically complete  but  is intensively used   in the context of  the Penrose limit \cite{b7}.
\par 
The situation becomes  more interesting if we take into account the conformal symmetry. First, let us recall that  the  maximal dimension of the conformal group of, non-flat,  metric $g$, given by (\ref{e000}),  is   seven \cite{b8a} (see also \cite{b8b,b8c}).  On the other hand, the metric $g$  admits a homothetic vector field. Thus, in  the non-flat case,  the metric $g$ with  $\mA_{1}$ or $\mA_{2} $ (and their suitable generalization to the  circular polarization)   exhibits the maximal,  seven-dimensional,  conformal symmetry. 
Further examples of the gravitational   plane waves carrying the maximal conformal symmetry entail a non-homothetic conformal vector field (which is rather rare, even in the case of  non-vacuum plane waves).  
In the case of  gravitational plane waves it turns out that,  besides  two mentioned profiles, there is only  one  metric family    when  the dimension of the conformal group is { seven}  \cite{b9a,b9b}  (the dimension of the isometry group is { five} and   there is a  proper conformal transformation).
\par This special family of the gravitational plane waves is given, in the case of  the linear polarization,  by the  metric (\ref{e00})  with the profile 
\be
\label{e12}
\mA (u)=\frac{c}{(u^2+\alpha u+\beta)^2},
\ee
In this case  the geodesic equations can be also analytically solved what enables us to  explicitly discuss interaction of a classical particle with gravitational plane waves.  In order to describe  the regular  gravitational pulse one should consider   (\ref{e12}) with nonsingular denominator (then  the metric is geodesically complete).
\subsection{Complete conformal case and  its  Dirac delta  limit}
Let us consider the profile  (\ref{e12})  of the form 
\be
\label{e10a}
\mA (u)=\frac 2 \pi\frac{\epsilon^3}{(u^2+\epsilon ^2)^2}.
\ee
Such a  choice  is dictated not only by demanding  the completeness of geodesics  but also  by the fact that the profile   (\ref{e10a}) can be used  for the study  of the  Dirac delta  pulse, namely 
\be 
\label{e1}
\lim_{\epsilon \to 0} \mA (u)=\delta (u).
\ee
In view of the above  we start with  $\epsilon<\pi$, i.e., sufficiently narrow gravitational pulse.  Then  the general solution of the geodesic equation takes the form:
\be
 X^\sigma(u)=C_1^\sigma \sqrt{u^2+\epsilon^2}\sin(a_\sigma \arctan(\frac {u}{\epsilon})+C_2^\sigma),
\ee
or equivalently 
\be
\label{e5}
X^\sigma (u)=D_1^\sigma \sqrt{u^2+\epsilon^2}\sin(a_\sigma \arctan(\frac {u}{\epsilon}))+D_2^\sigma \sqrt{u^2+\epsilon^2}\cos(a_\sigma \arctan(\frac {u}{\epsilon})),
\ee
and 
\be
V(u)= C_4+C_3u+\frac14\sum_{\sigma}(C_1^\sigma)^2\left[u\cos(2 a_\sigma\arctan(\frac u \epsilon)+2C_2^\sigma)-\epsilon a_\sigma \sin (2a_\sigma \arctan(\frac u \epsilon)+2C_2^\sigma)\right],
\ee
or equivalently
\be
\label{e7}
\begin{split}
V(u)&= C_4+C_3u+\frac14\sum_{\sigma}((D_1^\sigma)^2-(D_2^\sigma)^2)\left[u\cos(2 a_\sigma\arctan(\frac u \epsilon))-\epsilon a_\sigma \sin (2a_\sigma \arctan(\frac u \epsilon))\right]\\
&-\frac12\sum_{\sigma}D_1^\sigma D_2^\sigma\left[u\sin(2 a_\sigma\arctan(\frac u \epsilon))+\epsilon a_\sigma \cos (2a_\sigma \arctan(\frac u \epsilon))\right],
\end{split}
\ee
where 
\be
a_\sigma =\sqrt{1-\frac {\sigma\epsilon}{\pi}}, \quad \sqrt{2}>a_->1, \quad 1>a_+>0.
\ee
\par 
Of course, one can consider the case $\epsilon \geq \pi$;  then  the $X^-$  solution takes the same form, but  in $X^+$ (and consequently  partially in $V$)   one should replace $\sin$ and $\cos$  by hyperbolic or linear functions; the slowly varying case should   coincide  with the  attractive (repulsive) character of the harmonic oscillator.  However, such a choice of $\epsilon$  can lead not only to the technical modification  -- see the discussion at the end of  Section \ref{s2}. Thus  in what follows we consider the  case $\epsilon<\pi$ which  seems more adequate  to analyse  the burst-like pulse  of gravitational waves.
\par
First, we  impose the initial  conditions.  Let us note that  the condition  $\dot X^\sigma (-\infty )=0$\footnote{Dot  refers to derivative with respect to $u$.}   implies $C^\sigma_2 =\frac \pi 2 a_\sigma$  and, consequently,  we may  define
\be
 X^\sigma_{in} =\lim_{u \to -\infty}X^\sigma(u) .
 \ee
 It is worth to notice that $ X^\sigma_{in} =\frac{a_\sigma}{\sin(a_\sigma \frac{\pi}{2})}X^\sigma(0)$.
  In consequence, the solution is of the form 
\be
\label{e13}
X^\sigma(u)=\frac{X^\sigma_{in}}{{\epsilon a_\sigma}}{\sqrt{u^2+\epsilon^2}}\sin(a_\sigma (\arctan(\frac u \epsilon )+\frac \pi 2)) .
\ee
The asymptotic behaviour near the  future infinity is
\be
X^\sigma(u)\approx X^\sigma_{in}\left(\frac{u}{\epsilon a_\sigma}\sin(a_\sigma\pi)-\cos(a_\sigma\pi)\right), \quad u\gg1 ,
\ee
 thus the relative transverse-space distance between two trajectories $X^\sigma_1, X^\sigma_2$ grows  linearly near the future infinity, so they exhibit the  so called  velocity memory effect \cite{b1d,b1e,b2c,b2d,b10} 
\be
\sqrt{\sum_{\sigma}(\dot X^\sigma_1(u)-\dot X^\sigma_2(u))^2}\approx \sqrt{\sum_{\sigma}( X^\sigma_{in1}- X^\sigma_{in 2})^2\frac{\sin^2(a_\sigma\pi)}{\epsilon^2a_\sigma^2}}, \quad u\gg 1.
\ee
\par 
In general,  we define the initial conditions as follows 
\be 
\label{e13a}
\dot X^\sigma _{in}\equiv \dot X^\sigma (-\infty )=\frac{P^\sigma_{in}}{P_V}\quad,     \quad X^\sigma_0\equiv X^\sigma(0);
\ee
then 
\be
\label{e2}
D_1^\sigma =C_1^\sigma\cos(C_2^\sigma)=\frac{1}{\sin(\frac \pi2 a_\sigma )}\left(X^\sigma_0\frac{\cos(\frac \pi 2a_\sigma)}{\epsilon}+\dot X^\sigma_{in}\right),
\ee
\be
\label{e3}
D_2^\sigma =C_1^\sigma\sin(C_2^\sigma)=\frac {X^\sigma_0}{\epsilon}.
\ee
Moreover, for a  timelike geodesic, i.e. $P_\mu P^\mu=-m^2$, one gets 
\be
\label{e3a}
2\dot V_{in}\equiv 2\dot V(-\infty)=\frac{-1}{P_V^2}(m^2+\sum_{\sigma}(P_{in}^\sigma)^2)<0.
\ee
Denoting $V(0)\equiv V_0$ one finds  
\be
\label{e4}
C_3=\dot V_{in}-\frac 14\sum_\sigma\left[((D_1^\sigma)^2-(D_2^\sigma)^2)\cos(\pi a_\sigma)+2D_1^\sigma D_2^\sigma\sin(\pi a_\sigma)\right] ,
\ee
and 
\be
C_4=V_0+\frac 12 \sum_\sigma \epsilon a_\sigma D_1^\sigma D_2^\sigma ,
\ee
where $D_1^\sigma$ and $D^2_\sigma $ are given by (\ref{e2}) and (\ref{e3}).
Thus we obtain the explicit form of the geodesics, in particular, the final velocities or momenta.
\par 
Due to  eq. (\ref{e1})  and eqs.  (\ref{e2})-(\ref{e4})  one gets, after some troublesome computations, the geodesic for the metric with  the Dirac delta profile 
\be
\label{e11}
\lim_{\epsilon \to 0}X^\sigma(u)=X^\sigma_0(1+\frac \sigma 2 u\theta (u))+\dot X^\sigma _{in }u ,
\ee
\be
\lim_{\epsilon \to 0}V(u)=C_4+\dot V_{in} u-\frac  14 \theta(u)\sum_\sigma\sigma (X_0^\sigma)^2-\frac 14 u\theta (u)\sum_\sigma\left( \frac{(X^\sigma _0)^2}{2}+2\sigma X^\sigma _0\dot X^\sigma_{in}\right)-\frac 12\sum_\sigma X^\sigma _0\dot X^\sigma_{in}, 
\ee
where $\theta(u)$ is  the Heaviside step function. 
Now  taking arbitrary $u_0<0$ one can find the value $C_4$ for the Dirac delta function, 
$C_4=V(u_0) -\dot V_{in}u_0+\frac 12 \sum_\sigma X_0^\sigma   \dot X^\sigma_{in}$ and, consequently, the final form of the  geodesics; they agree with the form  presented in the literature, see  e.g.,  \cite{b2f, b11}; this confirms the conclusion that  the impulsive limit is totally independent of the special form of the original profile \cite{b5h}. 
\subsection{Conformal symmetry for the Dirac delta profile}
As we have  indicated above, the metric $g$  with the profile (\ref{e10a}) exhibits the conformal symmetry \cite{b9a,b9b}.  The  generators are   strictly related to the decomposition of $X^\sigma$ in terms of the initial conditions (cf. (\ref{e5}),  (\ref{e2}) and (\ref{e3}))
\be
X^\sigma(u)=X_0^\sigma P_1^\sigma(u)+\dot X^\sigma_{in} P_2^\sigma(u) .
\ee
The Killing vectors are  defined as follows 
\be
\begin{split}
\hat V&=\partial_V,\\
\hat D_1^\sigma&=P_1^\sigma(u)\partial _\sigma-X^\sigma\dot P^\sigma _1(u)\partial_V ,\quad \\
\hat D_2^\sigma&=P_2^\sigma(u)\partial _\sigma-X^\sigma\dot P^\sigma _2(u)\partial_V.\quad
\end{split}
\ee
The nonvanishing commutators are:
\be
[\hat D_1^\sigma, \hat D_2^\sigma]=-\frac{a_\sigma}{\sin(\frac \pi 2 a_\sigma)}\hat V.
\ee
Next, there is a standard homothetic generator 
\be
\hat H =2V\partial_V+X^+\partial_{X^+}+X^-\partial_{X^-},
\ee
and the proper conformal one 
\be
\hat K=u^2\partial _u-\frac 12((X^+)^2+(X^-)^2)\partial _v+uX^+\partial_{X^+}+uX^-\partial_{X^-}+\epsilon^2\partial_u.
\ee
They  satisfy the following commutations rules 
\be
\label{e6}
\begin{split}
[\hat H,\hat V]&=-2\hat V, \quad [\hat H,\hat D_1^\sigma]=-D_1^\sigma, \quad [\hat H,\hat D_2^\sigma]=-\hat D_2^\sigma,\\
[\hat K,\hat D_2^\sigma]&=\frac{a_\sigma\epsilon^2}{\sin(\frac \pi 2 a_\sigma)}\hat D_1^\sigma-\frac{\epsilon a_\sigma\cos(\frac \pi 2 a_\sigma)}{\sin(\frac \pi 2 a_\sigma)}\hat D^\sigma _2,\\
[\hat K,\hat D_1^\sigma]&=\frac{\epsilon a_\sigma\cos(\frac \pi 2 a_\sigma)}{\sin(\frac \pi 2 a_\sigma)}\hat D^\sigma _1-  \frac{a_\sigma}{\sin(\frac \pi 2 a_\sigma)}\hat D_2^\sigma.
\end{split}
\ee
Now,  taking the limit $\epsilon\rightarrow 0$,  one obtains for the Dirac delta profile the following  generators of the conformal   algebra 
\be
\begin{split}
\hat V&=\partial_V,\quad \hat H =2V\partial_V+X^+\partial_{X^+}+X^-\partial_{X^-},\\
\hat D_2^\sigma&=u\partial _\sigma-X^\sigma\partial_V, \quad \\
\hat D_1^\sigma&=\partial_\sigma+\frac{\sigma}{2}\theta(u)\hat D_2^\sigma,\\
\hat K&=u^2\partial _u-\frac 12((X^+)^2+(X^-)^2)\partial _v+uX^+\partial_{X^+}+uX^-\partial_{X^-},
\end{split}
\ee
satisfying commutation relations
\be
\begin{split}
[\hat H,\hat V]&=-2\hat V, \quad[\hat D_1^\sigma,\hat D^2_\sigma]=-\hat V,  \quad [\hat H,\hat D_1^\sigma]=-D_1^\sigma, \\
\quad [\hat H,\hat D_2^\sigma]&=-\hat D_2^\sigma,  \quad [\hat K,\hat D_1^\sigma]=-\hat D_2^\sigma,
\end{split}
\ee
due to $u^2\delta(u)=0$ or  taking $\epsilon \rightarrow 0$  in (\ref{e6}).  This yields the conformal algebra of the metric with the Dirac delta profile and   generalizes to this case the  results for   the isometry algebra considered in Refs.  \cite{b2f,b12a,b12b}. 
\subsection{The cross section and the  change of energy}
Having explicit form of geodesics   one can compute the classical differential  scattering cross section associated  with the transverse-space scattering  map  in terms of the outgoing momentum  components (see \cite{b11} for more details).
\be
d \sigma_{classical}=dX^+_{in}dX^{-}_{in}=|J|dP^+_{out}dP^-_{out},
\ee
where $J$ denotes the Jacobian of the  transformation  between $P^\sigma_{out}$ and $X^\sigma_{in}$. 
Indeed,  for the profile (\ref{e10a}) with $\epsilon<\pi$, by virtue of (\ref{e9}) and (\ref{e13}) one gets
\be
\label{e15}
P^\sigma_{out}=P_V\dot X^\sigma_{out}=\frac{P_V\sin \pi a_\sigma}{\epsilon a_\sigma}X^\sigma_{in},
\ee
 thus 
\be
|J|=\frac{-\epsilon^2a_+a_-}{P_V^2\sin(\pi  a_-)\sin(\pi  a_+) }.
\ee
Taking the limit $\epsilon \rightarrow 0$ one obtains 
\be
|J|=\frac{4}{P_V^2},
\ee
 and, consequently,  the classical cross section for the Dirac delta profile.
\par 
In view of the above  it is also   tempting to compute the change of energy of a test particle  after the pulse has passed.  Below  we present  a  superficial approach  based on the   global  B coordinates; however, one should keep in mind that   such considerations  call for  more physical clarification  related to the choice of the inertial frame after and before the pulse, measurement problem and etc. 
To this end    let us recall some basic relations between the   light-cone and Minkowski approaches to the  relativistic particle with constant  four-velocity. 
In our  case, i.e., $\mA =0$,   not only $P_V$   is constant but also $P_\sigma=P^\sigma= \dot { X}^\sigma P_V$ (cf. (\ref{e14}) and (\ref{e9})). Thus, by virtue of (\ref{e9a}), (\ref{e9b}), (\ref{e13a}) and (\ref{e3a}), in this case   the geodesic is of  the  form
\be
\begin{split}
X^\sigma(u)&=\frac{ P^\sigma}{P_V}u +X^\sigma_0,\\
V(u)&=-\frac{m^2+\sum_\sigma P_\sigma P^\sigma}{2P_V^2}u+V_0.
\end{split}
\ee 
Introducing $Z,T$ coordinates  as follows
\be
\label{e8}
u=\frac{Z-T}{\sqrt 2}, \quad V=\frac{Z+T}{\sqrt 2}, 
\ee
one obtains the  Minkowski metric with the signature   $(-,+,+,+)$  as well as 
\be
\begin{split}
T(u)&=\frac{1}{\sqrt 2}\left[-(1+\frac{m^2+\sum_\sigma P_\sigma P^\sigma}{2P_V^2})u+V_0\right],\\
Z(u)&=\frac{1}{\sqrt 2}\left[(1-\frac{m^2+\sum_\sigma P_\sigma P^\sigma}{2P_V^2})u+V_0\right].
\end{split}
\ee
Let us stress that, due to our convention, eq.  (\ref{e8}), time  runs in the opposite direction to the $u$ coordinate; in consequence, $u=\pm\infty$  corresponds to $T=\mp\infty$. Moreover, we assume $P_V <0$ in order to obtain the same direction for $\lambda$ and $T$, cf. (\ref{e9}). 
Denoting by 
\be
\gamma=\frac{-P_V}{\sqrt 2}(1+\frac{m^2+\sum_\sigma P_\sigma P^\sigma}{2P_V^2}),
\ee
we have
\be
P^\sigma=\gamma W^\sigma, \quad P_V=\frac {\gamma}{\sqrt 2}(W^Z-1),
\ee
where $W^\sigma, W^Z$ are    velocities in the  Cartesian coordinates.  Then   the energy of  a test particle can be expressed as follows
\be 
E=\frac{m}{\sqrt{1-\sum_\sigma W_\sigma W^\sigma -(W^Z)^2}}=\gamma=\frac{P_V}{\sqrt 2}(\dot V-1).
\ee
Now, we are in the position to analyse the change of energy after the wave has  passed. Namely,  due to (\ref{e7}) 
\be
\dot V_{out}=\dot V(u=\infty)=\dot V_{in}-\sum_\sigma D_1^\sigma D_2^\sigma \sin(\pi a_\sigma)=\dot V_{in}-2\sum_\sigma \frac{\cos(\frac \pi 2 a_\sigma)}{\epsilon}X_0^\sigma\left(\frac{\cos(\frac \pi 2 a_\sigma)}{\epsilon}X_0^\sigma+\dot X_{in}^\sigma\right).
\ee
Thus 
\be
\label{e10}
\bigtriangleup E=E(u=\infty)-E(u=-\infty)=-\sqrt{ 2} P_V \sum_\sigma \frac{\cos(\frac \pi 2 a_\sigma)}{\epsilon}X_0^\sigma(\frac{\cos(\frac \pi 2 a_\sigma)}{\epsilon}X_0^\sigma+\dot X_{in}^\sigma),
\ee
or equivalently
\begin{align}
\bigtriangleup E&= -\sqrt{ 2} P_V \sum_\sigma \frac{\cos(\frac \pi 2 a_\sigma)}{\epsilon}X_0^\sigma(-\frac{\cos(\frac \pi 2 a_\sigma)}{\epsilon}X_0^\sigma+\dot X_{out}^\sigma)\\
&=-\frac{P_V}{\sqrt 2}\sum_\sigma X_0^\sigma \dot  X_0^\sigma \frac{\sin(\pi a_\sigma)}{\epsilon a_\sigma}.
\end{align}
In order to analyse this result let us note that, in our convention,  $E(T=\infty)-E(T=-\infty)=-\bigtriangleup E$  and $\dot X^\sigma _{out}=\dot X^\sigma (u=\infty)=0$ is equivalent to  $\dot X^\sigma (T=-\infty)=0$;  thus  assuming  vanishing  transverse-space velocities in the past infinity, in particular the particle at rest,  we obtain that the    energy does not decrease   after the wave has passed
\be
E(T=\infty)-E(T=-\infty)=-\bigtriangleup E=-\sqrt{ 2} P_V \sum_\sigma \frac{\cos^2(\frac \pi 2 a_\sigma)}{\epsilon^2}(X_0^\sigma)^2 \geq 0.
\ee
In the general case, the final energy  is less, greater or  equal to  the initial energy, depending on the  relations  between initial positions and velocities (one can find the suitable conditions), cf.  results in Ref. \cite{b13}. However, as we mentioned above, further clarifications are necessary.
\par 
Finally, taking the limit $\epsilon\rightarrow 0$  in eq. (\ref{e10}) one  obtains  the change of energy in the case of the Dirac delta profile:
\be
\bigtriangleup E=-\frac{ P_V}{2\sqrt{2}} \sum_\sigma \left (\frac{(X_0^\sigma)^2}{4}+\sigma X_0^\sigma \dot X_{in}^\sigma\right ).
\ee
\section{Plane waves in  the Baldwin-Jeffery-Rosen coordinates}
\label{s2}
\subsection{General discussion}
In order to analyse the interaction  or scattering  one  should  specify  the notion of the  pulse of  gravitational waves.  First, it seems that the pulse cannot be  defined only by the vanishing of the wavefront in the past (future) infinity since the  asymptotic behaviour  of geodesic might be,  for example, oscillatory. Thus our starting point  is  the linear behaviour of  geodesics near   the past (future) infinity.   Let us consider the first set of geodesic equations (\ref{e9a}).  It turns out that by means of  classical results on differential equations \cite{b14},  the condition  $u^2\mA(u)\in L^1(\mR)$  is sufficient to ensure the  linear behaviour of the geodesic, $X^\sigma(u) \simeq  a^\sigma u+b^\sigma$ for large $|u|$;\footnote{Very often one can come across  slightly  weaker  integrability condition for   $u\mA(u)$,  instead of $u^2\mA(u)$, however, it implies only a finite limit $\frac{X^\sigma(u)}{u}$ as  $|u|$ tends to the infinity.} more intuitively, every integral curve has a unique
slant asymptote at $\pm\infty$, distinct integral curves having distinct asymptotes, and
every slant straight line is the asymptote of a unique integral curve (moreover, under the assumption that $\mA$ is of  constant sign for large $|u|$  this condition is also necessary).
In consequence, we have 
\be
\label{e41}
\lim_{u\mapsto \pm\infty} u^2\mA(u)=0.
\ee
Now, let us note that the differential equation for $V$ (\ref{e9b}) can be integrated to the form 
\be
\label{e42}
\dot V(u)=-\frac 14 \sum_\sigma \sigma (X^\sigma(u))^2\mA(u)-\frac 12 \sum _\sigma (\dot X^\sigma(u) )^2+\textrm{const },
\ee
which, by virtue of eq. (\ref{e41}),  implies the linear behaviour of $V$  for large $|u|$. 
\par 
So far we  discussed   gravitational plane waves in the  B coordinates where   both   the wave and the   geodesic are global with no singularity;  B coordinates   cover the whole  plane wave spacetime by a single chart. However, the gravitational plane waves   are frequently discussed  in the, so called,  Baldwin-Jeffery-Rosen   (BJR) coordinates  for which 
\be
\label{e43}
g=\sum_{\sigma,\sigma'}a_{\sigma\sigma'}(u)dx^\sigma d x^{\sigma'} +2dudv ,
\ee
where  $a(u) = (a_{\sigma\sigma'}(u))$ is  a positive matrix, see e.g.,  \cite{b1b,b1c,b2c,b2d}. The BJR coordinates, in contrast to the  B ones,  are not harmonic and typically
not global, exhibiting  $u$ coordinate singularities.  This fact is reflected  in transformation between   both  coordinates. Namely, only a piece of the  B manifold  can be covered by  the BJR coordinates  and consequently at least  two BJR  maps are needed to completely describe the interaction (scattering) of particle by gravitational plane waves. 
The definition of the  BJR coordinates    is  related to the geodesic equations (\ref{e9a}) and  consequently is not unambiguous. Thus we impose some physical conditions to specify them. Namely,  we will require  that the B and BJR coordinates coincide in the past and future infinity since there is no gravitational wave.
\par  The  next problem is the minimal number of charts to cover the whole B manifold. 
We  will give criteria such that  there exist two,  say "in" and "out",  BJR  charts which cover the whole $B$ manifold  and we will  express all information concerning   interaction   in terms of them. 
\par
To do this let us note that the linear asymptotic behaviour  implies that there are  solutions $P^\sigma_{out} (u)$ such that 
\be
\label{e44}
\lim_{u\mapsto \infty}P^\sigma_{out} (u)=1; 
\ee
moreover, due to Theorem 2 in  Ref. \cite{b14}, we have the following  estimate 
\be
\label{e45}
|P^\sigma_{out} (u)-1|\leq\exp(G(u))-1, \quad \textrm{ for } u>0,
\ee
where 
\be
\label{e46}
G(u)\equiv \frac 12 \int_u^\infty \tilde u|\sigma \mA(\tilde u)|d\tilde u=\frac 12\int_u^\infty \tilde u|\mA(\tilde u)|d\tilde u,
\ee
which gives $P^\sigma_{out} (u)>0$ for $u\geq0$ provided that
\be
\label{e46a}
G(u)<\ln(2) ,\quad  \textrm{ for }   u\geq 0.
\ee
Since $G(u)\leq G(0)$ for $u\geq0$ thus   $P^\sigma_{out} (u)>0$ for $u\geq 0$  if the inequality 
\be
\label{e47}
\int_0^\infty  u|\mA( u)|d u<2\ln(2),
\ee
holds. 
Thus (\ref{e47})   is a sufficient (but not necessary)   condition for  $P^\sigma_{out}(u)>0$ for $u\geq 0$.
On the other hand, applying standard reasoning (see, e.g., \cite{b2c,b11}) to the solutions  $P^+_{out},P^-_{out}$ one can find   $u_0>0$ such that $(P^+_{out}P^-_{out})(u)$ vanishes at $-u_0<0$\footnote{Some aspects of the behaviour of $P^\sigma$ appear also in  the study of caustic and focusing properties  of plane waves, see e.g., \cite{b11,b15a,b15b,b15c}}. 
 \par To simplify our further  considerations let us  assume that the profile $\mA$ is an even function,  $ \mA(-u)=\mA(u)$.
 Then  the functions $P^\sigma_{in} (u)\equiv P^\sigma_{out}(-u)$  are  also solutions of (\ref{e9a}) and satisfy the following conditions 
 \be
 \label{e48}
 \lim_{u\rightarrow -\infty }P_{in}^\sigma(u)=1, \quad P_{in}^\sigma (0) =P_{out}^\sigma (0) , \quad (P^+_{in}P^-_{in}) (u_0)=0.
 \ee
Let us now analyse  the linear independence of these solutions. 
As we  showed above,  at least one of the functions $P^\sigma_{out}$ vanishes somewhere,  say    $P^-_{out} (-u_0)=0$;    then $P^-_{in}$ and $P^-_{out}$ are linearly independent. Now, if $P^+_{out}(-u_0')=0$ for $u_0'>0$  then $P^+_{in}$ and $P^+_{out}$ are  also linearly independent (and we redefine  $u_0$  by $\min(u_0,u_0')$ in further considerations). If $P^+_{out}>0$ on the whole  real line   and  $\lim_{u\rightarrow -\infty} P^+_{out}(u)=1$ then  there exists, in one direction, a globally defined map ($x_{in}^+=x^+_{out}$ in further considerations); in  the other case $ P^+_{out}$ and $P^+_{in}$ are linearly independent.  Thus without loss of generality we may assume that the Wronskian 
 \be
 \label{e49}
 \mW^\sigma (u)\equiv \dot P^\sigma_{in}(u)P^\sigma_{out}(u)-\dot P^\sigma_{out}(u)P^\sigma_{in}(u)\equiv  \mW^\sigma=\textrm{const},
 \ee
 is not zero. 
 \par 
 In view of the above the condition (\ref{e47})  ensures that two BJR charts are sufficient (when the profile in an even function). Namely, we  introduce the BJR coordinate $(u,x_{in}^\sigma,v_{in})$   for  $u<u_0$     
 \begin{align}
 \label{e50}
 X^\sigma&=P^\sigma_{in}(u) x^\sigma_{in},\\
 \label{e51}
 V=&v_{in}-\frac 1 4 \sum_{\sigma}(x_{in}^\sigma )^2\dot a_{in}^\sigma(u) ,
 \end{align}
 where $a^\sigma_{in}(u)=(P^\sigma_{in}(u))^2 $. Then the metric  takes the form 
 \be
 \label{e52}
 g_{in}=2dudv_{in}+\sum_\sigma a^\sigma_{in}(d x_{in}^\sigma)^2 .
 \ee
 Similarly, by means of  the functions $P^\sigma_{out}$, we introduce the  BJR coordinates $(u,x_{out}^\sigma,v_{out})$ in the region   $u>-u_0$  leading to the metric $g_{out}$ in this region.  Let us stress  that the metrics (coordinates)   coincide with the B ones in the past  (future) infinity.  In the common domain $(-u_0,u_0)$  we have the transformation rules 
\begin{align}
\label{e53}
x^\sigma_{out}&=x^\sigma_{in}\frac { P_{in}^\sigma }{P_{out}^\sigma},\\
\label{e54}
v_{out}&=v_{in}-\sum_\sigma \frac{\mW^\sigma P_{in}^\sigma}{2P_{out}^\sigma}(x_{in}^\sigma)^2,
\end{align}
which  transform the metric $g_{out}$ into $g_{in}$. 
\par Now let us analyse the behaviour of the geodesics in BJR coordinates in  both charts. First, let us recall that the geodesics,   for $ u<u_0 $,  are of the form,  see e.g. \cite{b2d} 
\be
\begin{split}
\label{e55}
x^\sigma_{in}(u)&= b^\sigma_{in}H^\sigma_{in}(u)+c_{in}^\sigma, \\
v_{in}(u)&=-\frac 12\sum_\sigma (b_{in}^\sigma)^2H_{in}^\sigma(u)+e_{in}u+d_{in},
\end{split}
\ee
 where  $b^\sigma_{in},c^\sigma_{in}, e_{in},d_{in}$ are  some constants and 
\be
\label{e56a}
H^\sigma_{in}(u)=\int_0^u \frac{1}{(P^\sigma_{in}(\tilde u))^2} d\tilde u, \quad u<u_0, \quad H_{in}^\sigma(0)=0.
\ee
Similarly, in the region $ u>-u_0 $ we have  geodesics $ x^\sigma_{out}(u), v_{out}(u)$ with some initial parameters $b_{out}^\sigma,c_{out}^\sigma,d_{out},e_{out}$.
\par  Let us find the relations between "in" and "out" initial conditions and compare them with the ones obtained in the B coordinates. 
First, substituting $ x^\sigma_{out}(u), x^\sigma_{in}(u)$ into (\ref{e53})  one obtains after some computations 
\be
\label{e57}
c_{in}^\sigma=c_{out}^\sigma, \quad b_{out}^\sigma =c^\sigma_{in }\mW^\sigma+b_{in}^\sigma.
\ee
Next, substituting $v_{out}(u),v_{in}(u)$ into (\ref{e54}),  and using the identity  $\mW^\sigma H^\sigma_{in}=1-\frac {P^\sigma_{out}}{P^\sigma_{in}} $   one gets 
\be
\label{e58}
e_{in}=e_{out}, \quad d_{out}=d_{in}-\frac12 \sum_{\sigma}\mW^\sigma (c^\sigma_{in})^2.
\ee
Thus we  expressed  all parameters of  geodesics after the wave has passed in terms of the  initial ones and, consequently, the scattering process in terms of BJR coordinates. 
\par
Now we  compare these results with the ones obtained in    B coordinates. Intuitively,  since the B and BJR coordinates  coincide for $u \rightarrow \pm\infty$  and   $c_{in}^\sigma =x^\sigma_{in}(0)$ one  gets by virtue of  eqs. (\ref{e57}) 

\be
\label{e59}
\dot X^\sigma_{out}=\dot X^\sigma(\infty)=\dot X^\sigma(-\infty)+X^\sigma(0)\frac{ \mW^\sigma }{P_{in}^\sigma (0)}=\dot X^\sigma_{in}+X^\sigma_0\frac{ \mW^\sigma }{P_{in}^\sigma (0)}.
\ee
Of course, this formula can be directly confirmed in B coordinates. Indeed,  due to  (\ref{e41})  one obtains  
 \be
 \label{e60}
\lim_{u\rightarrow \infty}\dot P^\sigma_{out}(u)P^\sigma _{in}(u)=-\lim_{u\rightarrow \infty}\frac{ {\ddot P^\sigma_{out}(u) (P^\sigma _{in}(u)})^2}{\dot P^\sigma_{in}(u)}=\frac {-\sigma}{2}\lim_{u\rightarrow \infty} {{\mA(u)} (P^\sigma _{in}(u)})^2\frac{P^\sigma_{out}(u)}{\dot P^\sigma_{in}(u)}=0,
\ee
(if $P_{in}^\sigma $ tends to the infinity we use L'Hospital's rule, otherwise   the above limit  is immediately  zero)  thus, by virtue of (\ref{e44}),  one obtains 
\be
\label{e61}
\mW^\sigma= \lim_{u\rightarrow \infty}W^\sigma(u)=\lim_{u\rightarrow \infty}\dot P^\sigma_{in}(u)P^\sigma_{out}(u)=\dot P^\sigma_{in}(\infty).
\ee
On the other hand,  standard computations yield
\be
\label{e62}
\dot X^\sigma(\infty)=\dot X^\sigma(-\infty)+X^\sigma_0\frac{ \dot P_{in}^\sigma (\infty)}{P_{in}^\sigma (0)}.
\ee
In consequence, we get the formula (\ref{e59}).
\par 
Similarly,  the $V$ coordinate  and $v_{in},(v_{out})$ coincide in  past (future)  infinity; thus, cf.  (\ref{e55}) and (\ref{e56a}),   we have 
\be
\label{e63}
\dot V(-\infty)=\dot v_{in}(-\infty)=-\frac 12\sum_\sigma (b_{in}^\sigma)^2+e_{in}, \quad
\dot V(\infty)=\dot v_{out}(\infty)=-\frac 12\sum_\sigma (b_{out}^\sigma)^2+e_{out}.
\ee
Taking into account $e_{in}=e_{out}$ and (\ref{e57})   we get the jump of the $V$ velocity in terms of  $P_{in}^\sigma$ and $P^\sigma _{out}$ 
\be
\label{e64}
\dot V(\infty)=\dot V(-\infty)-\sum_{\sigma}\frac{\mW^\sigma}{P^\sigma_{in}(0)}X^\sigma(0)\dot X^\sigma(-\infty)-\frac{1}{2}\sum_{\sigma}(\mW^\sigma)^2\frac{(X^\sigma(0))^2}{(P_{in}^\sigma(0))^2},
\ee
which can be directly confirmed in terms of B coordinates (see (\ref{e42}) and (\ref{e59})).
\subsection{The conformal case and the  Dirac delta profile}
Let us now apply the above considerations to the  profile defined by eq.  (\ref{e10a}) with $\epsilon <\pi$. In this case 
\be
\label{e65}
\int_0^\infty  u|\mA(u)| du =\frac{ \epsilon}{ \pi}<2\ln 2,
\ee
thus  the sufficient condition (\ref{e47})  is satisfied and  we need only two  BJR  charts. Explicitly,  they are defined as follows 
\be
\label{e66}
\begin{split}
P^\sigma_{in}(u)&=\frac{\sqrt{u^2+\epsilon^2}}{\epsilon a_{\sigma}}\sin\left(a_{\sigma}\left(\arctan (\frac{u}{\epsilon})+\frac \pi 2\right)\right),\\
P^\sigma_{out}(u)&=P^\sigma_{in}(-u)=-\frac{\sqrt{u^2+\epsilon^2}}{\epsilon a_{\sigma}}\sin\left(a_{\sigma}\left(\arctan (\frac{u}{\epsilon})-\frac \pi 2\right)\right),
\end{split}
\ee
and satisfy the desired conditions, i.e., they tends to $1$ as $u\rightarrow \pm\infty$.  Moreover, $P^+_{in}(u)>0$ for $u\in\mR$ and $P^-_{in}$ vanishes only at one  point $u_0$
\be
\label{e68}
u_0=-\epsilon\cot(\frac {\pi}{a_-})>0;
\ee 
also
\be
\label{e69}
\mW^\sigma=\frac{\sin(\pi a_{\sigma})}{\epsilon a_{\sigma}}\neq 0.
\ee
In consequence,  we can define two maps, which overlap for $u\in(-u_0,u_0)$, and find the explicit form of geodesics  in terms of the BJR coordinates; in fact,  
\be
\label{e70}
H^\sigma_{in} (u)={\epsilon}{a_\sigma}\left [-\cot\left(a_\sigma\left (\arctan(\frac {u}{\epsilon})+\frac \pi 2\right)\right)+\cot\left(a_\sigma\frac \pi 2\right)\right],
\ee
and $H^\sigma_{out} (u)=-H^\sigma_{in} (-u)$ 	 (see eq.  (\ref{e55})).  Now, using  the relations (\ref{e59}) and (\ref{e64})  one can  confirm the  results obtained in Section \ref{s1}.   
\par
In this context  the case of the Dirac delta profile  seems to be especially interesting. Using the above results,  in the limit $\epsilon \rightarrow 0$, one obtains $u_0=2$ and  $\mW^\sigma=\frac{\sigma}{2}$ together with 
\be
H^\sigma_{out}(u)=\frac{u}{1-u\theta(-u)\frac{\sigma}{2}}, \quad {u>-2;} \quad H^\sigma_{in}(u)=\frac{u}{1+\frac{\sigma}{2}u\theta(u)  }, \quad u<2,
\ee
and, consequently, the geodesic for the Dirac delta profile (see also \cite{b2f,b11}).
For example, if  $\dot x_{in}^\sigma(-\infty)=0$ then $x^\sigma_{in}(u)=c_{in}^\sigma=x^\sigma_{in}(0)=x^\sigma_{out}(0)$ for  $u<2$. On the other hand,  due to (\ref{e57}), for $u>-2$   
\be
x^\sigma_{out}(u)=x^\sigma_{out}(0)+\frac{\sigma x^\sigma_{out}(0)}{2}H^\sigma_{out}(u)=\frac{1+\frac{\sigma}{2}u\theta(u)}{1-\frac{\sigma}{2}u\theta(-u)}x^\sigma_{out}(0),
\ee
and thus 
\be 
\dot x^\sigma_{in}(0)=0, \quad  \dot x^\sigma_{out}(0)=\frac{\sigma x^\sigma_{out}(0)}{2} =\frac{\sigma X^\sigma(0)}{2};
\ee
so  the jump  at zero the  velocity $\dot X^\sigma(u)$ in the  B coordinates, see  (\ref{e11}),    is encoded in  two maps in BJR coordinates; the same situation holds for the whole trajectory.
\par
Finally,  let  us briefly consider  the case of large $\epsilon$. As we mentioned in Section \ref{s1}, in this case the  form of the solution  $X^-(u) $  is  also valid. However,  for sufficiently large $\epsilon$ (i.e., $\epsilon\geq3\pi$),  by virtue  of  (\ref{e68}),     $u_0 \leq 0$. In consequence, we cannot cover the   whole $B$ manifold by two, "in" and "out",  BJR maps. Moreover,  for $\epsilon=(n^2-1)\pi$, $n=2,3,\ldots$ one obtains $a_-=n$; consequently, $X^-$, defined by (\ref{e13}), is a quotient of the  polynomial by	a  function and possesses $n-1$ zeros,  complicating in this way  the problem of  BJR  charts.  What is more,  by virtue of (\ref{e15}), $\dot X^-_{out}=0$.
Thus,  there   is no velocity memory effect in the $X^-$ direction; however, there is  also no permanent  displacement. Namely  taking  $\epsilon=(n^2-1)\pi $ in (\ref{e13}) one checks that 
\be
X^-_{out}=\lim_{u\rightarrow \infty}X^-(u)= (-1)^{n-1}X^-_{in}, 
\ee 
 thus  $ |X^-_{out1}-X^-_{out2}|=|X^-_{in1}-X^-_{in2}|$. For  example, in the  simplest case  $\epsilon=3\pi $, i.e., $n=2$  one gets
\be
X^-(u)=\frac{-u}{u^2+9\pi^2}X^-_{in},
\ee
and  $X^-_{out}=-X^-_{in}$.
\section{Summary and outlook}
We  discussed   analytically  some elements of the interaction of classical particles with  a pulse of  gravitational plane waves. The key point is the conformal symmetry of a certain class of plane waves metrics.  In particular, we confirm, directly by taking an appropriate limit, some results for  impulsive gravitational waves.  Furthermore, we gave certain conditions in order to  describe complete  interaction (scattering) in terms of BJR coordinates and  presented  an explicit illustration of  such situation. 
\par These results may provide  a starting point  for further considerations.  Let us  point out a few of them.\\
i) In the context of some optical effects in nonlinear gravitational plane waves  \cite{b2a,b16} one can consider massless particles following  null geodesics of the  metric (\ref{e00}) with the profile  (\ref{e12}).\\
ii) The generalization to case of polarized gravitational plane waves is also possible.   To this end let us note that the metric (\ref{e000}) with non-zero $\mA_\times$    also exhibits the  conformal symmetry \cite{b9a,b9b}. Thus it would be  interesting  to analyse some of the  recently obtained results \cite{b2g,b16b}   in this case.  \\
iii) It seems that the  problem of interaction of  a  quantum  particle  with exact gravitational plane waves (see \cite{b11}), especially the quantum cross sections, should be directly computable  in the case  of  the profile (\ref{e12}).\\
iv) The isometry group of gravitational plane waves can be identified with the so called the Carrol group (an ultrarelativistic group in 2+1 dimensions) without rotations \cite{b18a}. On the other hand, the  conformal extensions of the Carrol group  were classified \cite{b18b} (some of them  can be identified \cite{b18c} with the asymptotic symmetries in general relativity). Thus there is a  question concerning  the relations between  conformal  algebra discussed in this paper  and these conformal extensions. \\
v) There are  a number of  papers concerning the collision of  gravitational plane waves (see  \cite{b17a} and  references therein) in particular the ones with the Dirac delta profile \cite{b17b}.  It would be instructive to describe such a situation by means of the discussed metric, especially  in the case of Dirac delta profile. \\
vi) One of the main motivations  of this paper is an attempt to  give  a better  insight into some   problems   occurring in the infrared structure of gravity \cite{b19}.\\
\par
{\bf Acknowledgments}\\
The  authors would like to thank    Piotr  Kosi\'nski for  stimulating discussions as well as Peter  Horvathy and Gary  Gibbons  for useful remarks and suggestions.   Comments of Joanna Gonera, Cezary Gonera and Pawe\l   \ Ma\'slanka  are  also  acknowledged.
This   work  has   been partially   supported   by   the   grant  2016/23/B/ST2/00727  of  National  Science  Centre, Poland. 

\end{document}